# Can a ferroelectric diode be a selector-less, universal, non-volatile memory?

Soumya Sarkar[1]*, Xiwen Liu[2]*, Deep Jariwala[3]*

[1]School of Electronics and Computer Science, University of Southampton, Southampton, United Kingdom

[2]Thrust of Microelectronics, The Hong Kong University of Science and Technology (Guangzhou), Guangdong, China

[3]Electrical & Systems Engineering, University of Pennsylvania, Philadelphia, PA, USA

* soumya.sarkar@soton.ac.uk, xiwenliu@hkust-gz.edu.cn, dmj@seas.upenn.edu

**Abstract.** Recent advances in silicon foundry-process compatible ferroelectric (FE) thin films have reinvigorated interest in FE-based non-volatile memory (NVM) devices. Ferroelectric diodes (FeDs) are two-terminal NVM devices exhibiting rectifying current-voltage hysteretic characteristics that enable self-selecting designs critical for high-density memory. We examine progress in FeDs based on CMOS-compatible HZO, AlScN, and emerging van der Waals ferroelectrics. While FeDs demonstrate promising ON/OFF ratios and rectification capabilities, they face persistent challenges including limited write-cycling endurance, elevated operating voltages, and insufficient read currents. We provide materials-focused strategies to enhance reliability and performance of FeDs for energy-efficient electronic memory applications, with emphasis on their unique self-rectifying capabilities that eliminate the need for selector elements in crossbar arrays for compute in memory applications.

**Introduction.**

The energy consumption of Artificial Intelligence (AI) computing models presents significant sustainability challenges, with over 50% of this energy expended in memory access operations driven by data movement between computing and memory hardware.[1] While CMOS-compatible flash memory serves as the dominant non-volatile memory (NVM) technology, its microsecond-range access times lag behind main memory's nanosecond-range operation, limiting data retrieval speeds and increasing power consumption.[2,3] Vertical hetero-integration of logic and memory to reduce separation distances to micrometers (µm) could enable high-bandwidth data transfer in a near-memory computing architecture beneficial for data-intensive operations.[4,5]

Crossbar arrays of two-terminal resistive NVMs represent a promising approach for data-intensive near-memory computing.[5] Ferroelectrics (FEs) offer ideal attributes for resistive NVM due to their bistable crystal structure ensuring non-volatility, non-destructive readout, and long data retention, while their ultrafast electric-field driven switching (<1 ns) minimizes write energy.[6-10] Despite recognition of their suitability for memory applications since the 1950s, conventional inorganic perovskite FEs such as Barium Titanate $BaTiO_3$ (BTO), Strontium Bismuth Tantalate $SrBi_2Ta_2O_9$ (SBT) and Lead Zirconium Titanate $Pb(Zr,Ti)O_3$ (PZT) have proven difficult to integrate with semiconductor processes due to volatile and highly diffusive elements. The recent discovery of CMOS-compatible FEs, including $Hf_xZr_{(1-x)}O_2$ (HZO), AlScN, and van der Waals (vdW) FEs, has revitalized interest in FE-based NVM devices.[11-15]

FE-based NVM devices can be classified into three categories based on readout mechanisms: (i) Ferroelectric random-access memory (FeRAM), relying on direct readout of switched polarization charge as a current pulse; (ii) Ferroelectric field effect transistors (FeFETs), which read polarization coupled with channel resistance; and (iii) Ferroelectric resistive memory, which reads current flowing through a FE layer. While the first two categories appear on major semiconductor technology roadmaps and in industry pipelines,[16] resistive memories based on CMOS-compatible ultrathin FEs such as ferroelectric tunnel junctions (FTJs) and ferroelectric diodes (FeDs) remain emerging technologies requiring further development.

This perspective focuses on FeDs, highlighting their progress and challenges for electronic memory development. FeDs share a two-terminal design with FTJs, offering high-density integration potential in crossbar arrays, but fundamentally differ in their carrier transport mechanisms and circuit function. We examine key materials challenges in FeD development and benchmark their performance against state-of-the-art resistive memories. We discuss strategies to enhance performance and reliability, including controlled doping to increase ON current, defect reduction to improve endurance, and engineering clean metal-FE interfaces to enhance ON/OFF and rectification ratios. Notably, we highlight FeDs' unique self-rectifying feature, which eliminates the need for additional selector elements in crossbar arrays – crucial for preventing unwanted sneak-path leakage currents – enabling higher-density memory architectures.

**What is a ferrodiode?**

The simplest structure for a FeDs is similar to a FTJ – a Metal-FE-Metal (MFM) configuration, and its operation is based on the concept proposed by Esaki in 1971.[17] Device switching relies on the 'electroresistance effect' where the height of the potential barrier between the two metal electrodes depends on the direction of the FE polarization, modulating the current flowing through the barrier.[18] Electroresistance is defined as the ratio of the current flow in the two opposite FE polarisation states. A key requirement for the electroresistance effect is asymmetry in the potential barrier profile. This asymmetry leads to uneven charge screening when FE polarisation is switched – resulting in barrier height modulation.[18] The asymmetry can be designed by selecting top and bottom metal electrodes (with different screening lengths), interface engineering via surface termination or the insertion of ultrathin spacer layers.[18]

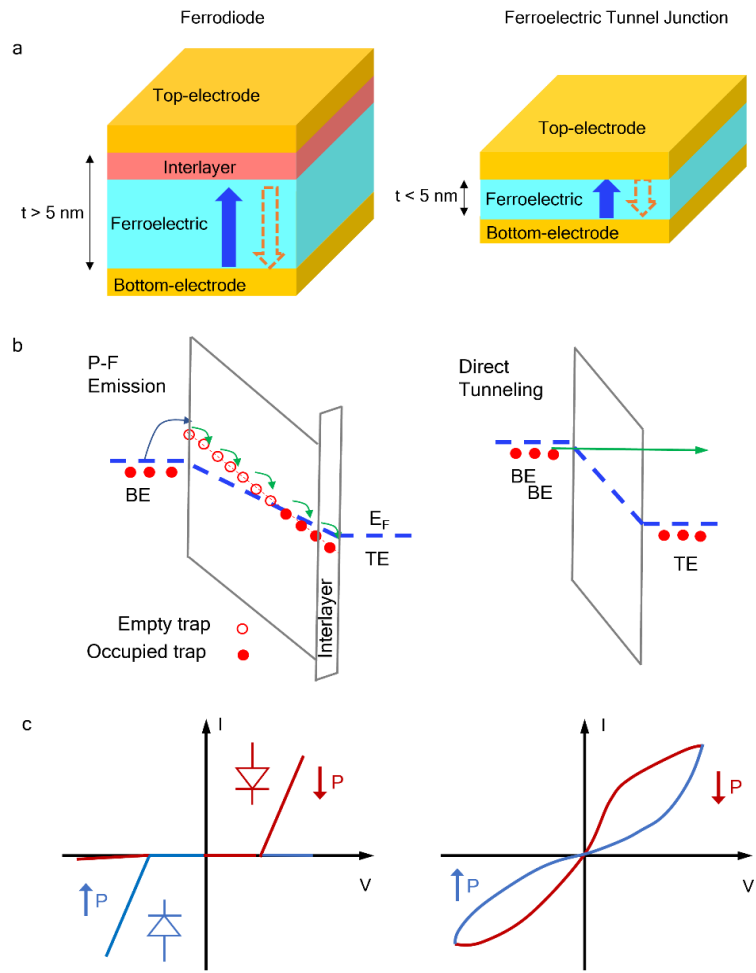

**Figure 1** Comparison of ferrodiodes (FeDs) with ferroelectric tunnel junctions (FTJs) in terms of (a) device structure, (b) energy level alignment and (c) current-voltage characteristics.

The fundamental difference between FTJs and FeDs lies in their transport mechanisms, which depend on the thickness of the FE layer. As illustrated in **Figure 1a** (right), the FE layer in an FTJ is ultrathin (< 5nm), allowing it to be fully carrier depleted. Charge transport occurs through direct quantum mechanical tunnelling (**Figure 1b**, right). The tunnel resistance depends exponentially on the square root of the barrier height, creating distinct ON and OFF states.[19] **Figure 1c** (right) describes the current-voltage (I-V) characteristics of an FTJ. Initially, an applied negative voltage aligns the net polarisation upward (FE is in OFF state). As a positive voltage is applied above the coercive field, the net polarisation switches downwards (FE is in ON state), leading to high tunnel current. The high ON current persists until a negative voltage that causes polarisation reversal is applied (FE is in OFF state), reducing the current flow.

In the case of FeDs, the FE layer is thicker (>5 nm), exceeding the 'direct tunnelling' limit. As a result, the charge transport is governed by polarisation dependent leakage mechanisms such as defect assisted tunnelling and Poole-Frenkel (PF) emission[20] (**Figure 1b**, left). Since the FE thickness in a FeD exceeds the depletion width, carrier transport is highly sensitive to the interface Schottky barrier, leading to a strong non-linearity and asymmetric diode like rectifying I-V characteristics. A choice of a suitable insulating interlayer that does not substantially affect the FE polarization can offer additional pathways to control the non-linear carrier transport (**Figure 1a**, left). **Figure 1c** (left) illustrates the I-V characteristics for a FeD. Initially, applying a negative voltage aligns the net polarization upward (FE is in OFF state). As the voltage sweeps positive under a forward bias, the current changes from low to high. Above the coercive field, the dipoles realign, switching the net polarization downward, and the diode polarity shifts from an OFF-forward diode to an ON-forward diode. The change in polarization from upward to downward direction represents the FeD switching from the OFF state to the ON state. When the voltage sweep direction is reversed toward negative voltages, the diode polarity reverses from an ON-reverse diode to an OFF-reverse diode. This strong non-linear I-V characteristics is a unique feature which precludes the need for an additional selector element (often a transistor) when a FeD is implemented in a crossbar architecture, reducing cell structure complexity from 1T1R to 1R (**Figure 2**).

Selector-less crossbar architectures with self-rectifying/non-linear FeD devices suppress sneak-path leakage by leveraging inherent device physics. The non-linear I-V characteristics (e.g., exponential threshold switching) ensure minimal off-state current when unselected cells experience sub-threshold voltages during read/write operations. Self-rectification introduces directional conductivity asymmetry, preventing reverse-bias leakage pathways while enabling forward-bias access to target cells – effectively isolating adjacent lines without requiring dedicated selector transistors. FeD crossbars theoretically enable ≥10X higher areal density than other NVM arrays by eliminating separate selector transistors and leveraging $4F^2$ cell sizes through vertical stacking of self-rectifying ferroelectric junctions.[21, 22]

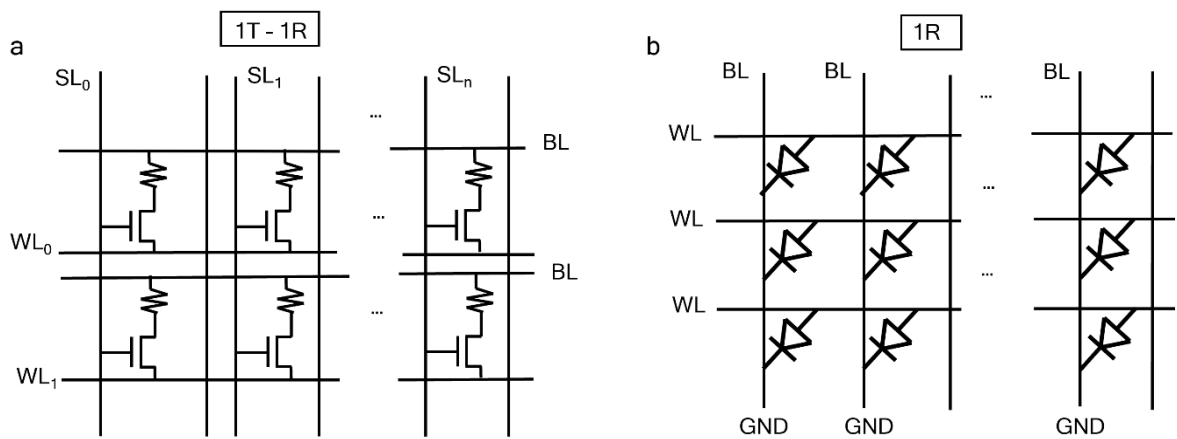

**Figure 2** A crossbar array with (a) two-terminal resistive NVM such as FTJs and RRAM and (b) FeDs. The self-rectifying nature of FeDs eliminates the needs of an additional selector element such as a FET, reducing the cell structure from 1T1R to 1R.

**Materials for FeDs**

FeDs were first demonstrated in inorganic perovskite oxides such as $BiFeO_3$ (BFO) and PZT in the 1990s.[23] However, challenges with CMOS-integration limited efforts to optimize FeD memory performance, despite their self-rectifying characteristics. Over the past five years, FeDs have gained renewed interest, driven by the rise of CMOS-compatible FEs. The field remains in its early stages and requires materials optimization for improved memory performance and reliability.

**Figure 3** highlights FeDs based on CMOS-compatible FEs, including HZO, AlScN, and vdW FE $CuInP_2S_6$ (CIPS). The FE properties of these materials have been discussed in previous review articles.[11-13] Briefly, FE in HZO arises from the stabilization of the non-centrosymmetric orthorhombic phase (**Figure 3a**), enabling switchable polarization of ~20 µC/cm² in films with thicknesses between 5–15 nm. A 10 nm thick HZO FeD with TiN top and bottom electrodes has demonstrated an ON/OFF ratio of ~$10^4$ and a rectification ratio of ~100 (**Figure 3d**).[21]

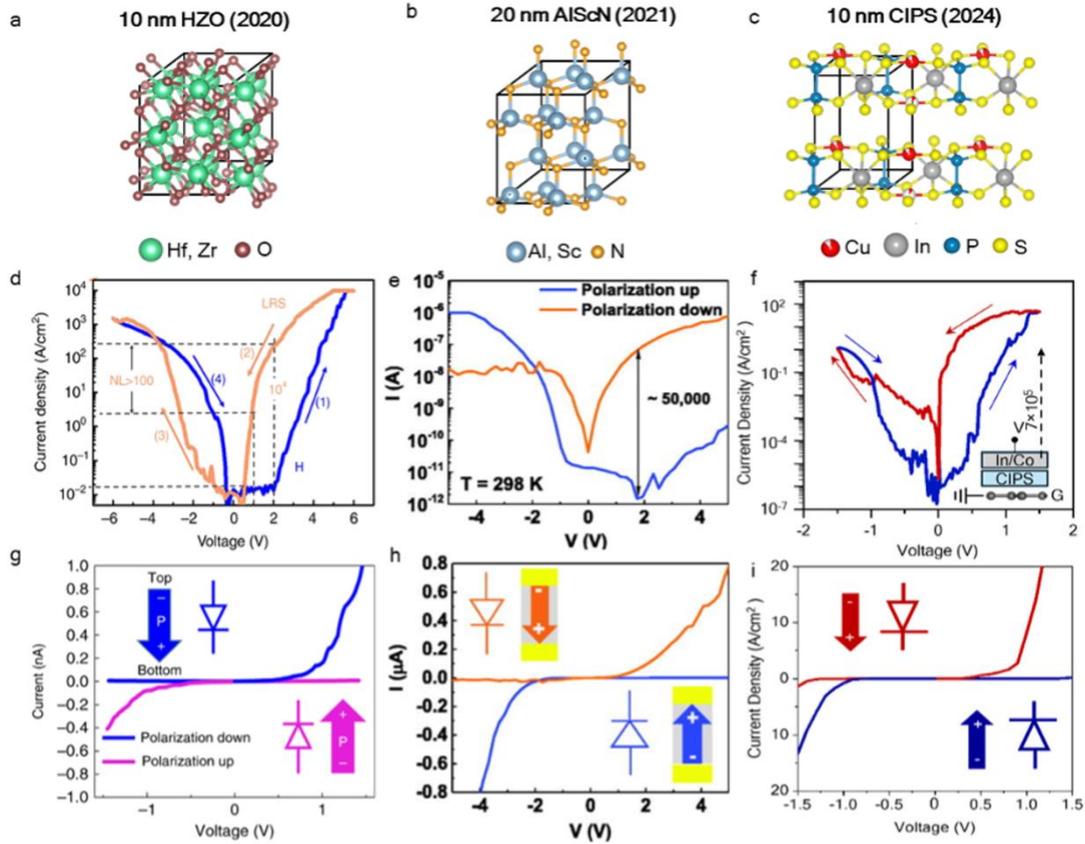

**Figure 3** (a-c) Models of HZO, AlScN, and vdW ferroelectric CIPS (d) Current density versus voltage characteristics for a 10 nm thick HZO FeD with TiN top and bottom contacts[21] (e) Current versus voltage characteristics of a 20 nm thick AlScN FeD with Pt top and bottom contacts. A thin native oxide interlayer present between the FE and the top contact introduces asymmetry in the average barrier height.[24] (f) Current density versus voltage characteristics of a CIPS FeD with graphene as a bottom contact and indium alloy van der Waals metallic top contact.[25] (g-i) The current-voltage characteristics plotted on a linear scale are representative of a FeD. The schematics in the inset depict the polarisation in the FE when the diode is forward/reverse biased.

The remarkable FE properties of wurtzite-structured AlScN (**Figure 3b**) were discovered in 2019.[26] AlScN exhibits a remanent polarization of ~50-100 µC/cm² and a coercive field of 2–4.5 MV/cm, making it suitable for electronic memory applications. 20 nm thick AlScN-based FeDs have achieved ON/OFF ratios of ~$10^5$ and rectification ratios of ~$10^3$ (**Figure 3e**) [24] The use of a native oxide interlayer (IL) between the AlScN layer and the top Pt metal contact introduces asymmetry in the Schottky barrier height, improving ON/OFF and rectification ratios beyond those of HZO-based FeDs.[27] Further, precise control over the IL thickness allows tuning of the interface depletion width and depolarizing field, enabling further modulation of electroresistance, switching voltages and FeD non-linearity.[27] AlScN ferroelectricity being extremely robust to temperatures, has also allowed the use of FeDs for high-temperature non-volatile memory, currently presenting itself as the most reliable and viable information storage mechanism in extreme temperature environments.[28, 29]

CIPS is a vdW ferrielectric (**Figure 3c**), where net polarization arises from the out-of-plane displacement of Cu and In ions in opposite directions within sulfur octahedra.[30] Compared to AlScN and HZO, CIPS exhibits a lower net polarization (~4 μC/cm²), reducing sensitivity to depolarization.[11] A robust ferroelectric response has been observed in films as thin as ~4 nm. Its layered structure and absence of dangling bonds enable FeDs based on vdW heterostructures. A graphene/CIPS/InCo FeD has demonstrated an ON/OFF ratio of ~$10^6$ and a rectification ratio of ~2500 (**Figure 3f**).[25] The high electroresistance and non-linearity arises from (i) ultraclean vdW contacts, which eliminate Fermi-level pinning allowing effective modulation of the interface Schottky barrier height upon FE polarisation switching, (ii) semi-metallic graphene bottom electrode, exhibiting low quantum capacitance near the Dirac point, and (iii) anisotropic electronic properties of vdW FEs, where the out-of-plane effective mass is three times higher than the in-plane direction in CIPS, enhancing carrier transport in the vertical direction.[31]

The current-voltage characteristics of HZO, AlScN and CIPS based FeDs have been plotted in a in linear scale in **Figure 3g, h, i** – similar to the characteristics in **Figure 1c** (left).

**Current status of FeDs**

While FeDs based on CMOS-compatible FE materials have demonstrated impressive device-level performance, further optimization is essential to address challenges related to scalability, read currents, data retention, and endurance. **Figure 4** benchmarks the performance of FeDs utilising CMOS-compatible FEs reported in literature, including perovskite FE oxides (BFO, PZT).[32-34] In addition, we compare the performance of alternative, technologically mature NVM, such as commercially fabricated RRAM from SkyWater and TSMC [35-37]

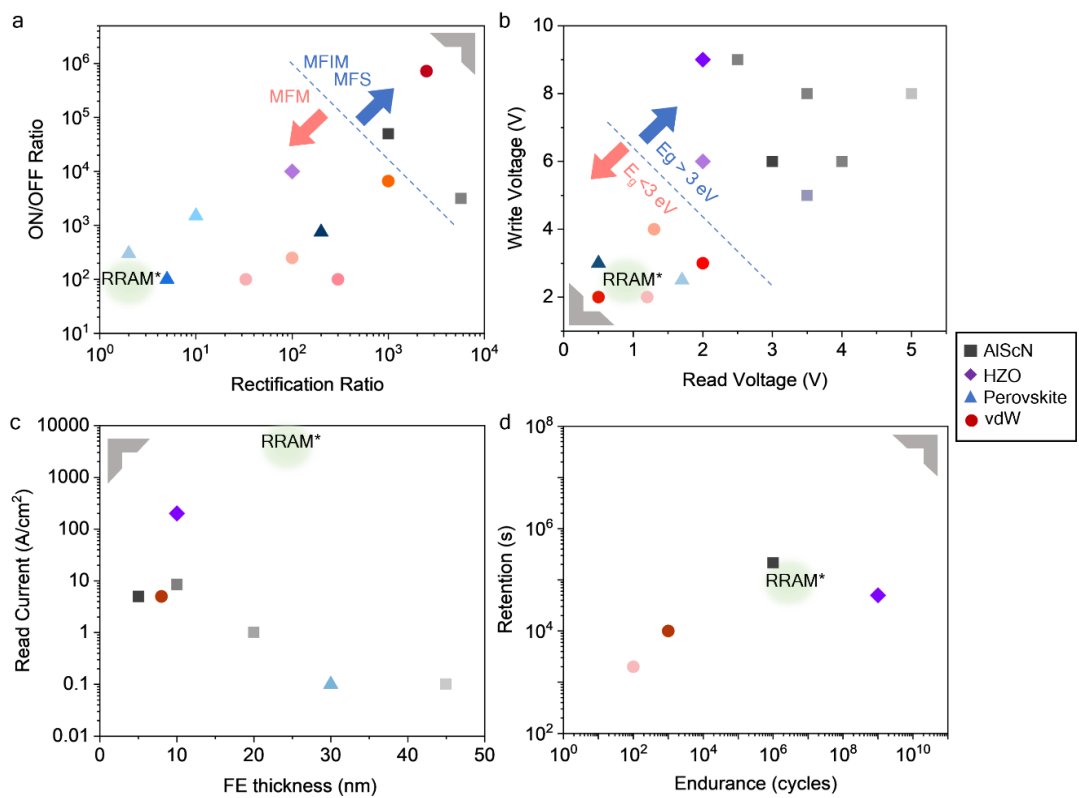

**Figure 4** Corner plots highlighting FeD performance reported in literature and its comparison with foundry processed (*) RRAM[35, 36] (green bubble – TSMC, SkyWater). (a) ON/OFF ratio vs Rectification ratio.[21, 24, 25, 30, 32-34, 38-42] (b) Read voltage vs write voltage[33, 38] (c) Read current vs FE film thickness, and (d) Data retention time vs memory endurance.[29] The grey arrow highlights the desired corner.

FeDs exhibit high ON/OFF ratios exceeding $10^5$ due to the electroresistance effect and low leakage currents in > 5 nm thick FE layers. In contrast, metal oxide-based commercial RRAM typically achieves ON/OFF ratios around 100 (**Figure 4a**). Moreover, FeDs exhibit current rectification ratios exceeding 5000, a unique device characteristic absent in most NVM devices.[21, 24, 30, 32-34, 38-42] Compared to FeDs with metallic contacts (MFM structures), those incorporating asymmetric metal contacts, insulating interlayers (MFIM structures), or semimetallic contacts (MFS structures) exhibit enhanced ON/OFF and rectification ratios. As discussed earlier, these high rectification ratios inherent to FeDs provide a self-selecting capability, a key advantage over alternative NVM technologies, which require additional selector circuits to mitigate sneak-path current leakage in crossbar architectures. The read/write voltages are similar or higher than RRAM (**Figure 4b**), which can be related to: (i) higher voltage drop across a thicker FE layer and (ii) the intrinsic FE coercive field. The dashed boundary in **Figure 4b** highlights a rough correlation between lower-bandgap ferroelectrics and reduced write/read voltages. vdW ferroelectrics (e.g., CIPS, $In_2Se_3$) have low bandgaps (<3 eV) and lower operating voltages. In contrast, high-bandgap, high-coercive-field materials such as HZO ($E_g$ > 5 eV) and AlScN ($E_g$ > 4 eV) exhibit higher read/write voltages for full polarization switching.

A persistent challenge for FeDs has been their low ON current, which limits miniaturization. Early demonstrations using perovskite oxide FEs exhibited ultralow read current densities (<1 A/cm$^2$). As shown in **Figure 4c**, most reported FeDs still operate below 10 A/cm$^2$ – insufficient for detection by peripheral circuits in memory chips.[21] In comparison, metal oxide based commercial RRAMs exhibit higher ON currents around $10^4$ A/cm$^2$. The low ON current in FeDs is related to the fundamental carrier transport mechanism, which relies on PF emission, and defect-assisted tunnelling. The large bandgap of FE materials results in low carrier density, reducing leakage current – a desirable feature for FeFETs and FeRAMs but a drawback for FeDs. The highest reported current density (~200 A/cm$^2$) has been observed in HZO-based FeDs, potentially due to mobile oxygen vacancies enhancing conduction.[21] Addressing this limitation will be crucial for FeD miniaturisation and integration with advanced CMOS nodes (<50 nm). It is expected that scaling FeDs to ~5 nm Fe thickness could lead to larger read currents enhancing their read speeds.

Endurance and retention are critical metrics for FeDs, yet available data remains limited (**Figure 4d**). The longest reported endurance without degradation has been demonstrated in HZO-based FeDs (~$10^9$ cycles), surpassing commercial foundry-processed RRAM (~$10^6$ cycles).[43] In terms of retention, AlScN-based FeDs have reported stable data storage for up to one million seconds,[29] and projected retention stability upto 10 years – comparable to alternative NVMs.

**The future of FeDs**

FeDs have made substantial progress over the past five years compared to early demonstrations in the 2000s. Moving forward, the community should focus on addressing the following challenges:

*Device Scalability*: Wafer scale growth of HZO and AlScN thin films with robust ferroelectric response on CMOS-compatible substrates has been demonstrated.[13] Future research should leverage this progress to fabricate FeD crossbar arrays, evaluate memory capacity, and assess device yield and variability. Reproducing current FeD performance in miniaturized devices (<100 nm cell size) is essential. However, for vdW FeDs, large-area growth remains a major hurdle. Current demonstrations rely on exfoliated thin flakes, ensuring good device performance but raising concerns about scalability.

*Response Time:* Response time is less of a concern for FeDs as FE switching is inherently ultrafast (< 1 ns), compared to charge migration-based NVM devices (>10 ns reponse time). HZO based FeDs have already demonstrated ~10 ns response time,[13, 21] and AlScN has been shown to switch at 60 ns.[44] The switching times of FeDs based on vdW FEs need to be estimated as most studies still rely on dc or ms pulsed measurements. Further, the ON/OFF and rectification ratios of FeDs should be assessed as a function of switching time with ultrafast (ns) pulsed measurements.

*Metal-Ferroelectric Interface:* A defect-free metal-ferroelectric interface is crucial for optimising ON/OFF and rectification ratios, as well as FeD endurance and retention. Interface defects lead to Fermi level pinning, limiting depletion width modulation and polarization switching-induced barrier height modulation. CIPS FeDs with ultraclean vdW metallic contacts have demonstrated superior ON/OFF and rectification ratios due to efficient barrier height modulation.[25] Additionally, interface defects can trigger the formation of conducting filamentary bridges, compromising endurance and retention.[45] Use of 1-2 nm thick insulator interlayers (IL) between the FE and metal contact has shown improvement in rectification ratios, as demonstrated in AlScN FeDs.[27]

*Carrier doping:* Low carrier density in conventional ferroelectrics results in low read currents, posing a challenge for scalability and integration. Since FeDs rely on leakage currents – often driven by defect-assisted tunnelling and Poole-Frenkel emission – controlled doping with vacancy defects could enhance conductivity and improve read currents. However, increasing carrier concentration must not compromise cycling endurance or device reliability. Raman and XPS analysis should be used to characterise doping levels and assess their impact on ferroelectric polarization and response times. Emerging low-bandgap semiconducting ferroelectrics such as 3R-$MoS_2$ could offer higher conductivity, which may further improve FeD read currents.[46]

*Density trade-offs*: A unique feature of FeD memory is the ability to partially program the ferroelectric layer to achieve intermediate resistance states between fully up and fully down polarisation states leading to multi-bit memory. While the exact mechanism that leads to multi-bit memory in FeDs remains unclear, is it expected to be constrained as FeD sizes and thickness scale down due to reduction in the number and size of polar domains. Use of interfacial strain or doping could be used to tailor polar domains in scaled FeDs and engineer robust multi-bit memory.

**Outlook**

Recent progress in FeDs with CMOS compatible FEs positions them as promising candidates for next-generation low-power, high-density non-volatile memory, offering self-selecting capabilities to simplify circuit design. To enable large scale adoption, key challenges must be addressed, including interface engineering with clean contacts, large area growth of vdW FEs, and improved read currents through doping and material innovations. Scaling FeDs below 100 nm diameters while maintaining performance (operating voltage, ON/OFF ratio, retention and endurance) is essential for integration with advanced CMOS nodes, unlocking applications in neuromorphic electronics and compute-in-memory architectures.

**Author contributions**

All the authors discussed the content and wrote the manuscript.

**Data availability**

The data that supports the findings of this study are available from the corresponding authors, upon reasonable request.

**Declarations**

The authors declare no competing interests.